\documentclass{procDDs}
\usepackage{color}

\title{Multidimensional tunneling between potential wells at non degenerate minima}

\author{Anatoly ANIKIN}
{Moscow Institute of Physics and Technology, Moscow, Russia}                 
{anikin83@inbox.ru}                                   
\coauthor{Michel ROULEUX}                       
{Aix Marseille Universit\'e, CNRS, CPT, UMR 7332, 13288 Marseille, France, Universit\'e de Toulon, CNRS, CPT, UMR 7332, 83957 La Garde, France}               
{rouleux@univ-tln.fr}                                               
\def\e{\mathop{\rm \varepsilon}\nolimits}
\begin {document}
\newcommand{\Lib}{{\rm Lib}}
\newtheorem{theo}{Theorem}
\newtheorem{prop}{Proposition}
\newtheorem{claim}{Claim}

\newtheorem{defin}{Definition}

\maketitle

\index{Author1, I.I.}                              
\index{Author2, I.I.}                              
\index{Coauthor, I.I.}                             %

\begin{abstract}
   We consider tunneling between symmetric wells for a 2-D
semi-classical Schr\"odinger operator for energies close to the quadratic
minimum of the potential $V$ in two cases: (1) excitations of the lowest frequency in the 
harmonic oscillator approximation of $V$;  (2) more general excited states from Diophantine tori
with comparable quantum numbers.
\end{abstract}

\section{Tunneling between double wells: a short review}

Tunneling for Schr\"odinger type operators involves various scenarios which depend on the details of the dynamics, ranging from
integrable or quasi-integrable systems, to ergodic or chaotic ones. 

Assume that $V$ is a smooth function,
symmetric with respect to $\{x_1=0\}$, and $\{V(x)\leq E\}$ consists in 2 connected components $(U_E)_{L/R}$ (the potential wells), while 
$\limsup_{|x|\to\infty}V>E$. We are interested in the semi-classical spectrum of
Schr\"odinger operator $P=-h^2\Delta+V$ on $L^2({\bf R}^2)$ near energy $E$, which consists in pairs $E^\pm(h)=E_k^\pm(h)$ exponentially 
close to eigenvalues 
$E(h)=E_k(h)$ 
of the Dirichlet realization of $P$ in some neighborhood of a single well. We will always assume (\cite{Ma},\cite{Ma2}) that $E(h)$ are simple 
(non degenerate) and asymptotically simple.
As a general rule, the energy shift $\Delta E(h)=E^+(h)-E^-(h)$ (or splitting of eigenvalues)
is related to so called Agmon distance $S(E)$ between the wells,
associated with the degenerate, conformal metric $ds^2=(V-E)_+dx^2$
that measures the life-span of the particle in the classically forbidden region $V(x)\geq E$. 
Much is known in the 1-D case, even for excited states, or in several dimensions for the lowest eigenvalues. 

At the higher level of generality, we only require that $V'(x)\neq0$ on $\{V=E\}=\partial U_L(E)\cup\partial U_R(E)$. 
In the 1-D case, Landau-Lifshitz formula reads
\begin{equation}
\label{LL}
\Delta E(h)=2{\omega h\over\pi}e^{-S(E)/h}(1+o(1))
\end{equation}
where $\omega={\partial p\over\partial I}$ is the frequency of the periodic orbit at energy $E$, and $2S(E)=I={2\pi}^{-1}\oint(E-V)_+\,dx$. 
In higher dimensions, the structure of the classical flow plays an essential r\^ole, so that we are left with the following equivalence
(see \cite{Ma2} for a precise statement): Assume $V$ is analytic. Then
the splitting $\Delta E(h)$ is non exponentially small with respect to Agmon distance
(i.e. for all $\varepsilon>0$, larger than a constant times $e^{-(S(E)+\varepsilon)/h}$, $0<h\leq h_\varepsilon$) 
iff the eigenfunctions of $P$, with eigenvalues $E^\pm(h)$,
are non exponentially small (i.e. for all $\e >0$, larger, in local $L^2$ norm, than a constant times
$e^{-\varepsilon/h}$, $0<h\leq h_\varepsilon$) 
in an open set where minimal geodesics, connecting the 2 wells, meet their boundary.
These propositions are true for instance when the flow is ergodic inside the wells, and false in case of separation
of variables (complete integrability). 

Here we are interested in the special case of ``tunnel cycles'' for quasi-integrable flows, for which propositions hold true.
Let $V$ have non degenerate minima $a_{L/R}$ with $V(a_{L/R})=0$,  
and $V_0=\sum_j\lambda_j^2z_j^2$, $\lambda_1<\lambda_2$ be the harmonic approximation (in local coordinates $z$)
around $a_{L/R}$ and $p_0(x, \xi)=\xi^2+V_0$, 
the quadratic part of $p(x,\xi)$ near 0. 

In 1-D the splitting between the lowest eigenvalues is found to be
\begin{equation}
\label{LowLy}
\Delta E(h)=2\sqrt{\pi\over e}{\omega h\over\pi}e^{-S_h/h}(1+o(1))
\end{equation}
$\omega=\lambda_1$ is the harmonic frequency, and $S_h$ half the action of the periodic orbit for the Hamiltonian 
with reversed potential $q=\xi^2-V$ at energy $-E$, $E=\omega h/2$. For higher energies we have
\begin{equation}
\label{ExLow}
\Delta E_m(h)=2b_m{\omega h\over\pi}e^{-S(E)/h}(1+o(1)),$$ where 
$$E=(2m+1)\omega h, \ b_m={\sqrt\pi(2m+1)^{m+1/2}\over2^mm!e^{m+1/2}}
\end{equation}
so long $mh\leq c$, $c>0$ small enough, which somehow ``interpolates'' between (\ref{LL}) and (\ref{LowLy}) since $b_m\to1$ as $m\to\infty$. 

In several dimensions, the splitting between the two lowest eigenvalues \cite{DA},\cite{A1},\cite{A2} is again of the form
$$\Delta E(h)=2\sqrt{\pi\over e}{\lambda_1 h\over\pi}e^{-S_h/h}(1+o(1))$$
Further, such formulas hold between  any
{\it low-lying eigenvalues}, i.e. for any $N$, there is $h_N>0$ such that for each principal quantum number $m\leq N$, the splitting $\Delta E_m(h)$ 
has an asymptotic of the form $\Delta E_m(h)\sim a_m(h)e^{-S_h/h}$ provided $0<h<h_N$
\cite{HeSj}, \cite{Ma}. See also \cite{MaRo} for degenerate minima. 

In this report we restrict our attention to KAM states, i.e. supported near a Diophantine torus and with quantum numbers $(k_1,k_2)$ 
such that $|k|h\leq c$, or semi-excited 
states in the limit $c\to0$, i.e. when $|k|\to\infty$ and $h\to0$ are related by $|k|h\leq h^\delta$, $0<\delta<1$. 
Further we shall only consider states (or approximate eigenfunctions) microlocalized on isotropic (generally Lagrangian) 
manifolds whose analytic continuation in the momentum space (i.e. in the classically forbidden region) are in a generic position. 
Lagrangian manifolds of 2 types are relevant to our analysis: (1) the flow-out of the boundary of the wells (2) the quasi-invariant tori
making a local fibration of the energy surface inside the wells. They have a (singular) limit as $E\to0$.

\section{Energy surfaces and librations}

The Lagrangian manifolds of the first type are the integral
manifold of $q$ passing above $(\partial U_E)_{L/R}$. 
From now on we assume that in local coordinates near $a_{L/R}$, $p(x,\xi)=p_0(x,\xi)+{\cal O}(|z|^3)$.
Consider first a single well $U_E$ then locally
$$\Lambda_\partial^E=\{\exp tH_q(\rho):\rho\in\partial U_E\times0, \ q(\rho)=-E, \ t\in{\bf R}\}$$
is a smooth real Lagrangian submanifold
of the form $\xi=\pm\nabla d_E(x)$, $x\notin U_E$, with a fold along $\partial U_E$. 
Here $d_E(x) = d_E(x,\partial U_E)$ is Agmon
distance from $x$ to $\partial U_E$ and satisfies (locally) the 
eikonal equation $\bigl(\nabla d_E(x)\bigr)^2=V(x)-E$. As $E\to0$, $\Lambda_\partial^E$ tends to the union
of the outgoing/incoming Lagrangian manifolds $\Lambda^\pm$ (called separatrices in 1-D) with a conical intersection at the origin.

We shall assume that $(\Lambda_\partial^E)_{L/R}$, as integral manifolds of Hamiltonian flow,  extend away from the wells as 
Lagrangian manifolds intersecting in the energy surface $\{q(\rho)=-E\}$ along a curve $\gamma_E$. This curve projects onto ${\bf R}^2_x$
precisely as a {\it libration} $\Lib _E$ between $U_L(E)$ and $U_R(E)$, i.e. a periodic orbit with end points at $\partial U_{L/R}(E)$ \cite{ArKoNe}.
We assume for simplicity there is exactly one such family of curves.
We call also $\Lib _E$ a {\it minimal geodesic} between $U_L(E)$ and $U_R(E)$ for Agmon distance $ds^2=\sqrt{(V(x)-E)_+}\,dx^2$. 
Assuming PT symmetry (i.e. $V$ symmetric with respect to $\{x_1=0\}$), we
denote by $\{x_E\}=\Lib _E\cap\{x_1=0\}$.
Then $d_E(x_E,U_L^E)=d_E(x_E,U_R^E)=S_E/2$, and 
$\Lib _E$ intersects $\{x_1=0\}$ at $x_E$ with a right angle.  
A neighborhood of $x_E$ in $\{x_1=0\}$ can be thought of as Poincar\'e section, intersecting $\gamma_E$ transversally.
The $\gamma_E$ are (unstable) periodic orbits of hyperbolic type, with real Floquet exponent $\beta(E)$. 
Of course, because of focal points, $(\Lambda_\partial^E)_{L/R}$ doesn't extend smoothly everywhere but only in a neighborhood of 
librations when the system is not integrable. 

As $E\to0$ the libration degenerates to an {\it instanton} $\gamma_0$. Parametrized as a bicharacteristic of $q(x,\xi)$ at $E=0$, it takes 
an infinite time to reach the equilibria $a_L$ or $a_R$ along $\gamma_0$. 
We shall assume that the stable outgoing and incoming manifolds $\Lambda^\pm_{L/R}$
at 0 intersect tranversally at $\gamma_0$. 

\section{Quasi-invariant Liouville tori}

Lagrangian manifolds of the second type are the invariant tori foliating (locally) the energy surface in the integrable case, or KAM tori, or
corresponding quasi-invariant tori in the quasi-integrable case. In the Section 6,
we shall also allow these Lagrangian manifolds to shrink to periodic orbits.

We can have already a good insight into the problem in replacing $V$ by its quadratic approximation.
This is what we call the {\it model case}. When
frequencies $\lambda_j$ are
rationally independent, we can essentially reduce to the model case by resorting to Birkhoff normal forms (or KAM theorem). 

So assume for simplicity that $p=p_0$ near $a_{L/R}$. Then for small $E>0$, the energy 
surfaces are foliated by invariant tori $\Lambda_\iota$, $E=2\lambda_1\iota_1+2\lambda_2\iota_2$
which can be extended in the complex domain along complex times, e.g. as
integral leaves $\widetilde\Lambda_\iota$ of $q(x,\xi)=\xi^2-\lambda_1^2 z_1^2-\lambda_2^2 z_2^2$, with purely imaginary time. 

The caustics of $\Lambda_\iota$ can be viewed as a rectangle shaped fold line 
delimiting the zone of pure
oscillations of the quasi-modes, and touching the boundary of the wells $\partial U_E$, $E=2\lambda_1\iota_1+2\lambda_2\iota_2$
at 4 vertices, the hyperbolic umbilic points (HU) points,
section of the torus by the plane $\xi=0$ in ${\bf R}^4$. We can identify $y$ with $\iota$. 
At the umbilic $y$, we have
$$T_y\widetilde \Lambda_\iota=T_y\Lambda_\iota=T_y(\hbox{fiber}), \quad T_y\Lambda_\iota\cap T_y\Lambda_\partial^E={\bf R}H_q,$$
where $E=2\lambda_1\iota_1+2\lambda_2\iota_2$.
More generally tori $\Lambda_\iota$ continue analytically in the $\xi$ variables as a 
multidimensional Riemann sheet structure, with a number of sheets 
corresponding to the choice of the sign of momentum, 
glued along the caustics, and all intersecting
at the HU's. 
On the other hand, $\Lambda_\partial^E$ has the fibre bundle structure
$\Lambda_\partial^E=\bigcup_{y \in \partial U_E} \gamma_y$
where $\gamma_y$ is the bicharacteristic of $q(x,\xi)$ at energy $-E$
issued from $\partial U_E$ at the point $y$. We have
\begin{equation}
\label{Gam_y}
\gamma_y=\widetilde \Lambda_\iota\cap \Lambda_\partial^E, \quad E=2\lambda_1\iota_1+2\lambda_2\iota_2
\end{equation}
with clean intersection.

Of course, in the general case (not model case), tori $\Lambda_\iota$ or $\widetilde \Lambda_\iota$
make only sense as asymptotic objects (via Birkhoff normal form) because they are not invariant under the
Hamilton vector flow. Assuming a Diophantine condition on $\lambda_1/\lambda_2$ 
we can also select a dense family of such invariant tori.

\section{The tunnel cycle and tunnel bicharacteristics}
\begin{figure}
 \centering
  \def\svgwidth{275pt}
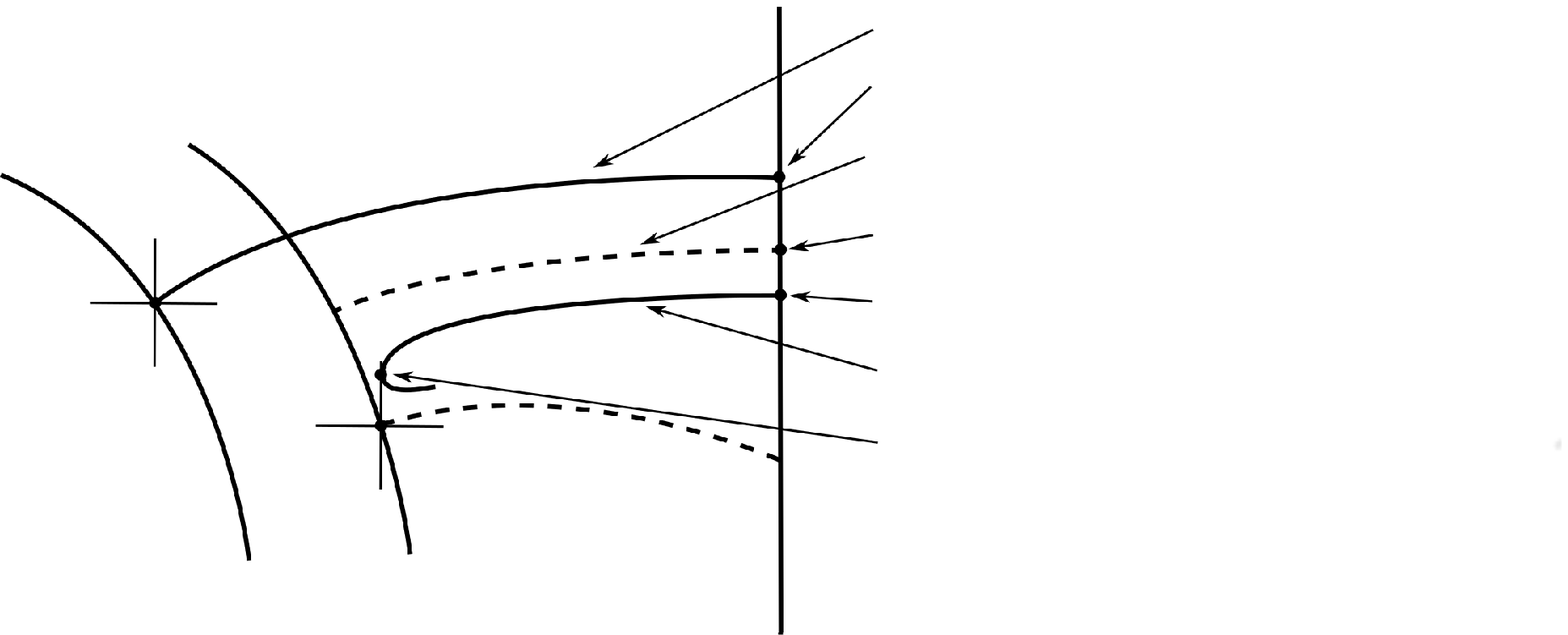
\label{fig1}
\end{figure}
If the system were integrable near 0, because of PT symmetry, the extension of 
$\bigl(\widetilde\Lambda_\iota\bigr)_{L}$ would usually coincide with $\bigl(\widetilde\Lambda_\iota\bigr)_{R}$, 
the decaying branch of $\bigl(\Lambda_\iota\bigr)_{R}$.
For a general, non integrable system, there is no reason for this holds and $\widetilde\Lambda_L$ intersects $\widetilde\Lambda_R$ 
along a one dimensional manifold.

\begin{defin}
Assume again there is only one libration $\Lib _E$. We call the lift $\gamma_E$ of $\Lib _E$ the {\rm tunnel cycle}.
We call the bicharacteristic $\widetilde\gamma\subset q^{-1}(-E)$ a {\rm tunnel bicharacteristic} 
if there are $\rho_L, \rho_R\in\widetilde\gamma$, with $E=2\lambda_1\iota_1+2\lambda_2\iota_2$ and 
$\rho_L\in\bigl(\widetilde\Lambda_\iota\bigr)_{L}$, $\rho_R\in\bigl(\widetilde\Lambda_\iota\bigr)_{R}$. We say also that $\rho_L, \rho_R$ are
{\rm in correspondence} along $\widetilde\gamma$.
\end{defin}
The tunnel cycle is a tunnel bicharacteristic for which $\rho_L, \rho_R$ are umbilics, but it carries generally no interaction between wells,
unless $\rho_L, \rho_R$ belong to quantized tori. But in a small, $h$-dependent neighborhood of $\gamma_E$ there are tunnel bicharacteristics
that carry interaction between wells (but generally do not close).
Non degeneracy of the tunnel cycle then implies the following:

\begin{prop}
Consider the model case. When $E=2\lambda_1\iota_1+2\lambda_2\iota_2$ we have
\begin{equation}
\label{Gam_E}
\gamma_E=(\Lambda^E_\partial)_L\cap(\Lambda^E_\partial)_R=\bigl(\widetilde\Lambda_\iota\bigr)_{L}\cap\bigl(\widetilde\Lambda_\iota\bigr)_{R}
\end{equation}
with a clean intersection.
\end{prop}
It follows from (\ref{Gam_y},\ref{Gam_E}) that along the tunnel cycle $\Lib _E$ we have simultaneously
$\gamma_E=(\Lambda^E_\partial)_L\cap(\Lambda^E_\partial)_R=\bigl(\widetilde\Lambda_\iota\bigr)_{L}\cap\bigl(\widetilde\Lambda_\iota\bigr)_{R}$
and $\gamma_E=(\widetilde\Lambda_\iota)_L\cap(\Lambda_\partial^E)_L=(\widetilde\Lambda_\iota)_R\cap(\Lambda_\partial^E)_R$
with clean intersections.

Unlike $\partial U_E$, the caustics of $\Lambda_\partial^E$ which is a smooth set, 
the caustics of $\widetilde\Lambda_\iota$ issued from $y$ is
a stratified set consisting of the umbilic $y$, and lines ${\cal C}_1(y)$, ${\cal C}_2(y)$ tangent at $y$ to the principal directions of $V''$. 
These caustics sets are the envelopes of Lissajous figures, whose lifts are (real) bicharacteristics of $q$.
Non degeneracy of the tunnel cycle $\gamma_E$ implies also the following splitting from (\ref{Gam_E})~:

\begin{claim} 
Let $\gamma_E$ be a minimal tunnel cycle, with end points $y^E_{L/R}$,
intersecting $\{x_1=0\}$ at $\Omega_E$, with $x_E=\pi(\Omega_E)$. For $y\in{\bf R}^2$ close to $y^E_{L/R}$, let $E(y)=V(y)$ and  
$\widetilde\Lambda_{\iota(y)}$ denote the Lagrangian manifold as above with HU $y$. 
Then for all $y$ close enough to $y^E_{L/R}$, we have:

1) $(\Lambda^{E(y)}_\partial)_L\cap(\Lambda^{E(y)}_\partial)_R$ is a curve $\gamma(y)$ whose projection is the libration $\Lib _{E(y)}$,
that intersects the caustics $\partial U_{E(y)}$ of $\Lambda^{E(y)}_\partial$ at some $y'(y)$ (both for L and R).

2) $\bigl(\widetilde\Lambda_{\iota(y)}\bigr)_{L}\cap\bigl(\widetilde\Lambda_{\iota(y)}\bigr)_{R}$ 
is a tunnel bicharacteristic $\widetilde\gamma(y)$, transverse to $\pi^{-1}(\{x_1=0\})$, 
$\widetilde\gamma(y)\cap\pi^{-1}(\{x_1=0\})=\{\widetilde\sigma(y)\}$,
and $\pi\bigl(\widetilde\gamma(y)\bigr)$ intersects orthogonally
$\{x_1=0\}$ at $\widetilde x(y)=\pi(\widetilde\sigma(y))$. 
Moreover $\widetilde\gamma(y)$ projects at some $\widetilde\rho(y)\in\Lambda_{\iota(y)}$ to $\widetilde y(y)$
tangentially to the caustics ${\cal C}(y)$ (both for L and R). 
\end{claim}
Thus $\gamma_E$, which
was common to both $(\Lambda^E_\partial)_L\cap(\Lambda^E_\partial)_R$
and $\bigl(\widetilde\Lambda_\iota\bigr)_{L}\cap\bigl(\widetilde\Lambda_\iota\bigr)_{R}$, 
splits into 2 distinct curves: (1) the lift of the libration at energy $E(y)$, 
(2) a tunnel bicharacteristic passing through the regular part of ${\cal C}(y)$.
Because the action along $\widetilde\gamma(y)$ gives the tunneling rate when $\Lambda_{\iota(y)}$ supports a quasi-mode we introduce the:

\begin{defin}
The action 
$\int_{\widetilde y(y_L)}^{\widetilde y(y_R)}\xi\,dx$ computed on $\widetilde\gamma(y)$ is called the {\rm tunnel distance} between
$\bigl(\Lambda_{\iota(y)}\bigr)_L$ and $\bigl(\Lambda_{\iota(y)}\bigr)_R$ (it equals Agmon distance when $\widetilde\gamma(y)=\gamma_E$.~)
\end{defin}

Let $y\in\partial_E(y)$. Integrating $\xi dx$ along $\gamma_y$ gives (locally) Agmon distance to the well~:
$$d_E(x)=\int_y^x\xi\,dx=\sum_j\lambda_j\int_{y_j}^{x_j}\sqrt{t^2-y_j^2}\,dt,\ x\in\gamma_y$$
Denote by $F_y^E(x)$ the RHS of this equation; provided $y\in\partial U_E$ is not too close to both $z$-axis, one can show that
$F_y^E(x)-d_E(x)$ is estimated by the square of the (Euclidean) distance of $x$ to its orthogonal projection on $\gamma_y$,
for $x$ in a neighborhood of $\Lib _E$. Similarly we consider variations from the regular part of the caustics ${\cal C}(y)$
$\inf\{\int_0^1\bigl(V(\gamma(s))-E\bigr)_+^{1/2}|\dot\gamma(s)|\, ds$, with
$\bigl(\gamma(0),\dot\gamma(0)\bigr)\in T{\cal C}(y), \gamma(1)=x$,
and write the critical value as
$G^E_{{\cal C}(y)}(x)=\int_{\widetilde y(x)}^x\xi\,dx$, or simply $G^E_{{\cal C}(y)}(x)=\int_{{\cal C}(y)}^x\xi\,dx$. Again 
$G^{E}_{{\cal C}(y)}(x)-d_E(x)+\int_{y}^{\widetilde y(x)}\xi\,dx=F_y^E(x)-d_E(x)$, where $\int_{y}^{\widetilde y(x)}\xi\,dx$,
$\widetilde y(x)\in{\cal C}(y)$ is a small error
term essentially independent of $x$ in a neighborhood of $\Lib _E$.

The next step consists in constructing quasi-modes. First we construct quasi-modes microlocalized on the $\Lambda_\iota$
selecting a sequence $\iota=\iota_k(h)$ from Bohr-Sommerfeld-Maslov (or EBK) quantization rules. 
As a rule, these (oscillating) quasi-modes extend in the shadow zone near $y_k(h)$ with exponential decay.
They can further be extended to $u_L$ and $u_R$
along $\widetilde\gamma(y_k(h))$ using WKB expansions, or the ``Gaussian beams'' method.
The eigenvalue splitting is given by the usual formula
\begin{equation}
\label{Spl1}
\Delta E_k(h)\sim 4h^2 \int_\Sigma u_L(0, x_2){\partial u_R \over \partial x_1}(0,x_2)\,dx_2
\end{equation}
where $\Sigma$ is a neighborhood of $x_E$ in $\{x_1=0\}$.
We now treat some specific cases in more detail.

\section{Tunneling near a pair of Diophantine tori}

Assume $c>0$ is so small that KAM theory ensures existence of a family invariant tori in the well $U_E=U_L(E)$  for $E\leq c$. 
We are interested in $\Delta E_k(y)$ for $E_k(h)$ near such fixed $E>0$. Assume that $\Lib _E$ starts at umbilic $y_E$ away from the $z$-axis,
and for simplicity, that
$y_E\in\Lambda_\iota$ with $\iota$ in the KAM set, i.e. such that the motion on $\Lambda_\iota$ is quasi-periodic with Diophantine frequency 
vector $\omega$ (this assumption seems to be generic, varying slightly $E$). 
In \cite{DoRo}, we proved the following~: Let 
$0<\delta<1$. Then in a $h^{\delta/2}$-neighborhood of $\Lambda_\iota$ in $T^*M$, there is 
a family $\Lambda_J$ of tori, labelled by their action variables $J=J_k(h)$ for $k\in{\bf Z}^d$ satisfying $|kh-\iota|\leq h^\delta$,
which verify Bohr-Sommerfeld-Maslov quantization condition, and are
quasi-invariant under $H_p$ with an accuracy ${\cal O}(h^\infty)$. 
At first approximation, the umbilics $y_k(h)\in\Lambda_J$ have the form
$y\sim(\lambda_1^{-1}\sqrt {2\lambda_1\iota_1}, 
\lambda_2^{-1}\sqrt {2\lambda_2\iota_2})$ or $y\sim(\lambda_1^{-1}\sqrt {2h\lambda_1k_1}, 
\lambda_2^{-1}\sqrt {2h\lambda_2k_2}), \ k=(k_1,k_2)=k(h)\in{\bf N}^2$
so the typical neighboring distance between $y_k(h)$ is $hE^{-1/2}$ when $y_E$ stays away from the $z$-axis.
Using Maslov canonical operator,
we obtain from these tori a sequence of quasi-modes for $P$ near $E$. By complex contour integrals (\cite{FuLaMa}, \cite{KaRo})) they extend
in a  $|h\log h|^{2/3}$- neighborhood of $U_E$, 
as states microlocalized on $\widetilde\Lambda_J$, and decaying
exponentially as $\exp[-F_y^E(x)/h]$, or $\exp[-G^E_{{\cal C}(y)}(x)/h]$. 
This decay propagates all along $\widetilde\gamma(y_k(h))$ and nearby bicharacteristics, which stay in the purely decaying branch
$\widetilde\Lambda_J$ of $\Lambda_J$.

Next we need to compare the tunnel distance with Agmon distance which coincide only on the tunnel cycle.
Let $S_L-S_R$ be the tunnel action between $y_L$ and $y_R$, we have at $\{x_1=0\}$ (see Fig.1)
\begin{equation}
\begin{aligned}
S_L&-S_R-2S_0(E)=2\bigl(F_{y}^{E(y)}(\widetilde x(y))-d_{E(y)}(\widetilde x(y))\bigr)\cr
&+2\bigl(d_{E(y)}(\widetilde x(y))-d_E(\widetilde x(y))\bigr)\cr
&+2\bigl(d_E(\widetilde x(y))-d_E(x_E)\bigr)\cr
\end{aligned}
\end{equation}
Evaluating each error term on the RHS, we arrive at $S_L-S_R-2S_0(E)=o(1)$, $h\to0$.
Then $S_L-S_R$ has a non degenerate critical point
at $\widetilde x(y_k(h))$ belonging to the tunnel bicharacteristic $\widetilde\gamma(y_k(h)$ common to $(\widetilde\Lambda_{J_k(h)})_L$ and
$(\widetilde\Lambda_{J_k(h)})_R$. 
The integral can be computed by standard stationary phase expansion around $x_k(h)$. Since the amplitude
of $u_R$ (and $u_L$) is non vanishing, we obtain eventually \cite{Creagh}
$$\Delta E_k(h)\sim B_k(h)e^{-(S_L-S_R)/h}$$
with $B_k(h)\sim{h^{3/2}\over\sqrt{\tau_L\sigma(H_L,H_R)\tau_R}}$.
Here $H_{L/R}$ are Hamilton vector fields tranverse to $\gamma_E$, and $\tau_{L/R}$ suitable Jacobians 
computed on $(\widetilde\Lambda_{J_k(h)})_{L/R}$.

\section{The quasi 1-D case}

In this section we shall assume that frequencies $\lambda_1,\lambda_2$ are non-resonant, with $2\lambda_1<\lambda_2$, and
the instanton $\gamma_0$ approaches the node singularity 
of the outgoing and incoming manifolds $\Lambda^{\pm}_{L/R}$ at $a_{L/R}$ in a regular direction (associated with $\lambda_1$).
We consider eigenstates with quantum vector $(m,0)$ for $m\in \mathbb{N}$, i.e.
$E_m=h(\lambda_1(2m+1)+\lambda_2)+{\cal O}(h^2)$, and compute
asymptotics for the energy splitting $\Delta E_m$ (as $h\to 0$, while $m$ stays fixed, and probably also when $hm\leq h^\delta$, $0<h<1$.) 
This amounts to let $\Lambda_\iota$ shrink to an isotropic torus.
\begin{theo}
\label{th_tp1}
Under the assumptions above
$$
\Delta E_m=2b_m\frac{\omega_1 h}{\pi} e^{-\frac{S(\widetilde E)}{h}}\big(1+o(1)\big),\quad h\to 0,
$$
where $b_m$ is found from (\ref{ExLow}), $S(\widetilde E)$ is half the action on $\Lib_{\widetilde E}$ at energy $\widetilde E=\widetilde E(h)$ 
which we determine as the solution of:
\begin{equation}
\label{e_of_h}
\tilde E+h\beta(\widetilde E)=h\Big(\lambda_1(1+2m)+\lambda_2\Big).
\end{equation}
Here $\beta(\widetilde E)$ is positive Floquet exponent of ${\rm Lib}_{\widetilde E}$.
\end{theo}
\par
In the case $m=0$ Theorem \ref{th_tp1} was proved, first, in \cite{D} when $\gamma_0$ is a straight line $x_2=0$, 
and then in \cite{A1},\cite{A2} in full generality (see also \cite{DA}). We want to show that passing to an arbitrary $m>0$ is quite simple.

{\it Sketch of proof}: We express (6) with the instanton phase ($E=0$).
The tunnel WKB approximation for the normalized quasimodes reads
$$
u_{L/R}=h^{-\frac{m+1}{2}}A_{L/R}(x)e^{-\frac{S_{L/R}}{h}}(1+{\cal O}(h)),
$$
where  $S_{L/R}=d_0(x,a_{L/R})$ (distance along the instanton), and the amplitudes $A_{L/R}$ are solution of the transport equation
\begin{equation}
\label{TrEq}
A\Big(\lambda_1(2m+1)+\lambda_2-\Delta S\Big)+2\nabla A\nabla S=0.
\end{equation}
Inserting it into (\ref{Spl1}) and applying asymptotic stationary phase, we obtain:
 \begin{align*}
\Delta E_m\sim 4h^{\frac{1}{2}-m}\sqrt{\pi}D^{-\frac{1}{2}}A_{L}^2(x_0)P_0e^{-\frac{S_{0}}{h}},
\end{align*}
where $x_0=x_E|_{E=0}$, $D=\frac{\partial^2S_L}{\partial x_2^2}(x_0)$; $P_0=\frac{\partial S_L}{\partial x_1}(x_0)$, and $S_0=2S_L(x_0)$.

From now on $\alpha\sim\beta$ means $\alpha=\beta(1+o(1))$ as $h\to 0$, and also we omit subscripts $L/R$.

To find $A(0)$ we shall solve the first transport equation (\ref{TrEq}) along the instanton $x=\gamma_0(t)$. Putting $b(t)=A(\gamma_0(t))$, we get
$b(0)=e^{-\omega_1 m t}\mathcal{J}(t)b(t)$, where
\begin{equation*}
\mathcal{J}(t)=
\exp\int_{0}^{t}\bigg(\frac{\Delta S}{2}-\frac{\lambda_1+\lambda_2}{2}\bigg)\,dt.
\end{equation*}
On the other hand, we can use harmonic oscillator approximation for $b(t)$ as $t\to \infty$. Therefore
$$b(t)\sim \frac{\sqrt[4]{\lambda_1^{1+2m}\lambda_2}2^{\frac{m}{2}}}{\sqrt{m!\pi}}\big(\xi_1(t)\big)^m,\quad t\to +\infty$$
where $\xi_1(t)$ is a $\xi_1$-coordinate of $\gamma_0(t)$.

\par
Defining
$\sigma =\lim_{t\to+\infty}e^{\lambda_1t}\xi_1(t)$ and $\mathcal{J}=\mathcal{J}(+\infty)$ we see that
\begin{equation}
\label{Spl2}
\Delta E_m\sim  \frac{2^{m+2}h^{\frac{1}{2}-m}}{m !\sqrt{\pi}D^{\frac{1}{2}}}\sqrt{\lambda_1^{2m+1}\lambda_2}\sigma^{2m}\mathcal{J}^2P_0e^{-\frac{S_0}{h}}.
\end{equation}

Let now $S_{E}$ be a half of the action along $\Lib_{E}$. In \cite{A1} we proved:
\begin{equation}
\label{ActAs}
S_{E}-S_0=\frac{E}{2\lambda_1}(1+\log 2)+E T_E+o(E),
\end{equation}
where $T_E$ stands for time to move along $\gamma_0$ between the intersections with $\partial U_E$. Inserting (\ref{ActAs}) 
with $E(h)=h(1+2m)\lambda_1$ into (\ref{Spl2}), we get
\begin{equation*}
\Delta E_m\sim \frac{2^{1-m}\sqrt{\pi}}{m! e^{\frac{1}{2}+m}}\frac{h\lambda_1}{\pi}\mathcal{T}\rho^{2m+1}e^{-\frac{S_E}{h}},
\end{equation*}
where
$$
\mathcal{T}=\mathcal{J}^2\frac{P_0}{\lambda_1\sigma}\frac{\sqrt{\lambda_2}}{\sqrt{D}},\qquad \rho=
\frac{\sigma\sqrt{\lambda_1}}{\sqrt{h}}e^{-\lambda_1 T_{E}}.
$$
One can easily see that
$
\rho\sim \sqrt{2m+1}
$, hence
\begin{equation}
\label{Spl3}
\Delta E_m\sim b_m\frac{h\omega_1}{\pi}\mathcal{T}e^{-\frac{S_{\varepsilon(h)}}{h}}.
\end{equation}
Thus, we arrived to the same formula as for $m=0$, but for the numerical factor $b_m$. 
The rest of proof is similar to the case $m=0$, its main ingredient is the following (see \cite{A2})
\begin{prop}
\label{Pr_Floquet}
$$
\beta(E)=\lambda_2-\frac{4\log \mathcal{T}}{T(E)}(1+o(1)),
$$
where $T(E)$ denote the period of $\Lib_E$.
\end{prop}
Note that proof of this Proposition uses assumption $2\lambda_1<\lambda_2$. When the instanton $\gamma_0$ is not a straight line, 
we resort to special coordinates (proposed in \cite{D4},\cite{D}): $s$ denotes arclength along $\gamma_0$, while $q$ is a coordinate 
along a normal to $\gamma_0$. But these coordinates are ill-behaved when Euclidean curvature of $\gamma_0$ tends to infinity near $a_{L/R}$,
which can happen, if $\frac{\lambda_2}{\lambda_1}\le 2$.


\section*{Acknowledgements}

The authors thank S. Dobrokhotov and J.Sj\"ostrand for their valuable comments.

\end {document}